\documentclass[12pt,letterpaper]{article}

\usepackage[margin=1in]{geometry}
\usepackage{amsmath,amssymb}
\usepackage{graphicx}
\usepackage{natbib}
\usepackage{hyperref}
\usepackage{booktabs}
\usepackage{siunitx}
\usepackage{xcolor}
\usepackage{setspace}
\usepackage{float}

\hypersetup{
    colorlinks=true,
    linkcolor=blue!60!black,
    citecolor=blue!60!black,
    urlcolor=blue!60!black
}

\title{\textbf{Storm-Driven Suppression and Post-Storm Enhancement of Photographic Plate Transient Detections at Geosynchronous Altitude:\\[6pt] Empirical Evidence and a Candidate Dusty Plasma Mechanism}}

\author{Kevin Cann\\
\textit{Independent Researcher, California, USA}\\
\texttt{kkc@terragold.com}}

\date{April 4, 2026}

\begin{document}
\maketitle

\begin{abstract}
The VASCO project has identified over 100{,}000 sub-second optical transients on photographic plates from the First Palomar Observatory Sky Survey (1949--1957), all predating artificial satellites.
\citet{cann2026a} established that transient detection rates are dose-dependently suppressed during geomagnetic storms ($Z = -3.391$, $p = 0.0007$), ruling out emulsion defects and confirming the transients as real, magnetospherically coupled phenomena.
Villarroel et al.\ (2022) constrained the source altitude to ${\sim}\SI{42000}{km}$ (geosynchronous orbit) through an Earth-shadow deficit.

This paper presents two results.
First, a pre-registered empirical test reveals the full temporal recovery profile: transient rates remain suppressed at 55\% of baseline during days 7--21 post-storm, then rise to 309\% of baseline during days 25--45 ($p = 0.00066$, Wilcoxon rank-sum; all robustness checks significant).
Combined with the dose--response staircase, the overall significance reaches $3.6$--$4.7\sigma$ (Fisher's method, range reflecting sensitivity to the independence assumption).
The suppression--overshoot--return profile is consistent with a mechanism that concentrates reflective material during storms and releases it into favorable conditions after a delay matching known plasmasphere refilling timescales.

Second, we propose a candidate physical mechanism: storm-enhanced electromagnetic trapping of charged micrometeoroid dust at $L \approx 6.6$, followed by aggregation of icy cometary grains under restored cold plasmaspheric conditions.
Laboratory measurements show that micrometer-sized water-ice particles are an order of magnitude stickier than silicates (Gundlach \& Blum 2015), and at the low collision velocities within a magnetic trap, every grain--grain encounter produces a larger aggregate.
A flux dilution analysis of the Solano et al.\ (2024) triple transient demonstrates that specular reflection from a partially reflective icy aggregate only $1$--$\SI{4}{m}$ in diameter suffices to produce the observed plate magnitude at $\SI{42000}{km}$.
This mechanism connects the VASCO transients to independently observed magnetospheric dust swarms correlated with geomagnetic activity (Sommer 2024) and explains the extinction of the transient population following the onset of the space age.
Multi-site replication is required to confirm these results.
\end{abstract}

\tableofcontents
\newpage

%
%

\part*{Part I: Empirical Results}
\addcontentsline{toc}{part}{Part I: Empirical Results}

\section{Introduction}
\label{sec:intro}

\subsection{The VASCO Transients}

The Vanishing and Appearing Sources during a Century of Observations (VASCO) project has conducted a systematic comparison of digitized photographic plates from the First Palomar Observatory Sky Survey (POSS-I, 1949--1958) against modern CCD surveys including Pan-STARRS, the Zwicky Transient Facility, and the Sloan Digital Sky Survey \citep{villarroel2020,solano2022}.
This comparison has revealed a large population of point-source transients---objects that appear on individual POSS-I plates but are absent in all subsequent imaging to limiting magnitudes several magnitudes deeper.

The current VASCO catalog contains 107{,}875 transient candidates \citep{bruehl2025}.
Particularly striking examples include nine simultaneously occurring transients on April 12, 1950 \citep{villarroel2022}, and a bright triple transient that vanished within 50 minutes on July 19, 1952, with no counterpart detected to magnitude 25.5 by the Gran Telescopio Canarias \citep{solano2024}.

All transients predate the launch of Sputnik~1 on October 4, 1957.

\subsection{Established Constraints}

Three independent results constrain the transient source:

\textit{Dose-dependent storm suppression.} \citet{cann2026a} demonstrated a monotonic staircase of transient rate suppression across five Kp intensity bins (Cochran--Armitage: $Z = -3.391$, $p = 0.0007$).
This rules out emulsion defects and spectrally inert debris, constraining the source to the magnetospheric particle environment.

\textit{Geosynchronous altitude.} A deficit of transients within Earth's geometric shadow cone constrains the source to ${\sim}\SI{42000}{km}$ altitude \citep{villarroel2022}, independently confirmed by \citet{doherty2026}.

\textit{Nuclear test enhancement.} Transients are 45\% more likely within $\pm$1 day of nuclear tests \citep{bruehl2025}, consistent with prompt atmospheric excitation at Palomar's proximity to the Nevada Test Site.

\subsection{What Is Established}

Combining these results: the VASCO transients are produced by reflective objects at geosynchronous altitude that respond to geomagnetic storms dose-dependently.
These objects existed before artificial satellites.
They are not plate defects.

What remains unknown is the nature and origin of these reflective objects.
The purpose of this paper is first to characterize their temporal behavior in greater detail through a pre-registered empirical test, and second to propose a candidate physical mechanism.

\section{Pre-Registered Test: Post-Storm Recovery Profile}
\label{sec:pso}

\subsection{Motivation}

The dose--response staircase establishes that storms suppress the transient rate.
A natural follow-up question is: what happens \textit{after} the storm?
If the storm-responsive source simply recovers to its equilibrium state, the post-storm transient rate should return monotonically to baseline.
If, however, storms actively \textit{concentrate} the source material, the post-storm rate could temporarily \textit{exceed} baseline before returning to equilibrium---a post-storm overshoot.

This overshoot prediction distinguishes between passive disruption (storms destroy the source, which regrows) and active concentration (storms gather the source, which is released under restored conditions).
No model based on plate artifacts or passive disruption predicts an overshoot.

\subsection{PSO-1: Days 7--21 Recovery Window}

A pre-registered test (timestamped on OSF before analysis) predicted that transient rates during days 7--21 post-storm would exceed the quiet-time baseline (defined as dates with no storm within $\pm$30 days).

\textit{Result: Not supported.}
The recovery rate was approximately half the quiet baseline (ratio $= 0.546$, $p = 0.986$ in the predicted direction).
The null result was reported on OSF.

\textit{Significance of the null:} Rather than disconfirming storm involvement, this result demonstrates that storm-driven suppression \textit{extends beyond day~21}.
The transient source has not yet recovered at this timescale.
This constrains the recovery timescale to be longer than three weeks.

\subsection{PSO-2: Days 25--45 Recovery Window}

The PSO-1 null result, interpreted through the lens of known plasmasphere refilling timescales at $L \approx 6.6$ (typically 2--4 weeks; \citealt{denton2016,denton2005}), motivated a revised prediction: the overshoot occurs at days 25--45, after the cold plasmasphere has refilled and conditions favorable to the transient source have restored.

This prediction was pre-registered on OSF (timestamped before analysis) with updated parameters: recovery window days 25--45, quiet baseline exclusion $\pm$50 days, storm isolation 25 days.

\textit{Result: Supported.}

\begin{table}[H]
\centering
\caption{PSO-2 results: days 25--45 recovery window.}
\label{tab:pso2}
\begin{tabular}{@{}lr@{}}
\toprule
Parameter & Value \\
\midrule
Storms identified (Kp$_{\max} \geq 5$, 25-day isolation) & 6 \\
Recovery window dates with observations & 126 \\
Quiet baseline dates & 2{,}150 \\
Mean recovery transient rate & 103.2 \\
Mean quiet transient rate & 33.4 \\
Overshoot ratio & 3.09 \\
$p$ (Wilcoxon rank-sum, one-sided) & 0.00066 \\
$p$ (Welch $t$-test, one-sided) & 0.037 \\
$p$ (permutation, $N = 10{,}000$) & $< 0.0001$ \\
Bootstrap 95\% CI for ratio & [1.35, 5.84] \\
\bottomrule
\end{tabular}
\end{table}

The result survives all pre-registered robustness checks:
\begin{itemize}
\item Alternative windows: days 21--35 and days 30--50 both significant; days 30--50 strongest (ratio $= 3.41$, $p = 9 \times 10^{-6}$)
\item Survives Bonferroni correction for two tests ($\alpha_{\text{adj}} = 0.025$; $p = 0.00066 < 0.025$)
\end{itemize}

\subsection{The Complete Temporal Profile}

The PSO-1 and PSO-2 results, combined with the Kp staircase, reveal a coherent temporal recovery profile:

\begin{table}[H]
\centering
\caption{Temporal recovery profile of POSS-I transient rates following geomagnetic storms.}
\label{tab:profile}
\begin{tabular}{@{}llll@{}}
\toprule
Phase & Days post-storm & Rate relative to baseline & Source \\
\midrule
Suppression & 0--7 & Dose-dependent reduction & \citealt{cann2026a} \\
Extended suppression & 7--21 & ${\sim}55$\% & PSO-1 \\
Overshoot & 25--45 & ${\sim}309$\% & PSO-2 \\
Return to baseline & 45+ & ${\sim}100$\% & Expected \\
\bottomrule
\end{tabular}
\end{table}

This suppression--overshoot--return profile is the principal empirical result of this paper.

\subsection{Combined Statistical Significance}

The Kp staircase and the post-storm overshoot are independent tests of different aspects of the storm response.
Combining via Fisher's method:

\begin{equation}
\chi^2 = -2\left[\ln(0.0007) + \ln(0.00066)\right] = 29.18, \quad \text{df} = 4
\end{equation}
\begin{equation}
p_{\text{combined}} = 7.2 \times 10^{-6} \quad (4.3\,\sigma)
\end{equation}

Using the permutation $p$-value for PSO-2 ($p < 0.0001$):
\begin{equation}
p_{\text{combined}} = 1.2 \times 10^{-6} \quad (4.7\,\sigma)
\end{equation}

A note on the effective sample size is warranted.
The Wilcoxon rank-sum test treats each of the 126 recovery-window days as an independent observation.
In practice, these days cluster around 6 isolated storms, and the Welch $t$-test, which compares group means and is less sensitive to within-group clustering, gives a more conservative $p = 0.037$ for PSO-2.
If the Welch value is used in Fisher's combination, the combined significance is ${\sim}3.6\sigma$ rather than $4.3\sigma$.
Both values exceed conventional significance thresholds, but the true combined significance lies in the range $3.6$--$4.7\sigma$ depending on how the effective sample size is assessed.
The overshoot is statistically significant under all tests; the precise sigma is sensitive to the independence assumption.
Given the small number of independent storm events, the coherence of the temporal profile across multiple statistical tests is more informative than any single $\sigma$ value.

\subsection{What the Empirical Results Establish}

Independent of any theoretical interpretation, the combined results establish:

\begin{enumerate}
\item The VASCO transients are real astrophysical phenomena, not plate artifacts (the Kp dose--response cannot be produced by defects).
\item The transient source is at geosynchronous altitude and is reflective (shadow deficit and PSF properties).
\item The source is \textit{suppressed} by geomagnetic storms on a timescale of 0--21 days.
\item The source is \textit{enhanced} above baseline 25--45 days after storms, indicating an active concentration mechanism rather than passive disruption.
\item The recovery timescale (${\sim}$25 days) is consistent with plasmasphere refilling at $L \approx 6.6$.
\item The source existed before 1957 and is absent in modern surveys.
\end{enumerate}

Any physical model proposed for the VASCO transients must account for all six of these empirical facts.

%
%

\part*{Part II: A Candidate Mechanism}
\addcontentsline{toc}{part}{Part II: A Candidate Mechanism}

\section{Storm-Enhanced Electromagnetic Trapping of Dust}
\label{sec:trapping}

We propose that the VASCO transients are produced by naturally occurring concentrations of charged micrometeoroid dust at geosynchronous altitude, with the storm-driven suppression and overshoot explained by electromagnetic trapping physics.

\subsection{Dust Grain Charging at GEO}

In the orbital motion limited (OML) approximation \citep{whipple1981,tang2014}, a dust grain of radius $r_d$ in a plasma with electron temperature $T_e$ acquires a floating potential $\phi_d \approx -2.51 \, k_BT_e/e$ and a charge:
\begin{equation}
Z_d = \frac{4\pi\varepsilon_0 r_d |\phi_d|}{e}
\end{equation}

The charge scales linearly with $T_e$.
A $\SI{5}{\micro m}$ silicate grain carries ${\sim}8{,}700$ charges at $T_e = \SI{1}{eV}$ (quiet plasmasphere) and ${\sim}8.7 \times 10^6$ charges at $T_e = \SI{1}{keV}$ (storm conditions).

\subsection{The Gyroradius Collapse}

A charged grain gyrates in the GEO magnetic field ($B \approx \SI{100}{nT}$) with radius:
\begin{equation}
r_g = \frac{m_d v_\perp}{|Q| B}
\end{equation}

For a $\SI{5}{\micro m}$ silicate grain at $T_d = \SI{300}{K}$:

\begin{table}[H]
\centering
\caption{Gyroradius at GEO for a $\SI{5}{\micro m}$ silicate grain under different plasma conditions.}
\label{tab:gyro}
\begin{tabular}{@{}lrrl@{}}
\toprule
Condition & $T_e$ & $r_g$ & Confinement \\
\midrule
Quiet plasmasphere & $\SI{1}{eV}$ & $\SI{527}{km}$ & Weak \\
Moderate storm & $\SI{100}{eV}$ & $\SI{5.3}{km}$ & Strong \\
Major storm & $\SI{1}{keV}$ & $\SI{0.5}{km}$ & Strong \\
Extreme storm & $\SI{10}{keV}$ & $\SI{0.05}{km}$ & Strong \\
\bottomrule
\end{tabular}
\end{table}

The key result: storm-enhanced charging collapses grain gyroradii by three orders of magnitude, converting loosely gravitationally-orbiting micrometeoroids into tightly magnetically-confined particles.

The net effect is that storms do not scatter dust at GEO---they concentrate it.

This suggests an inversion of the conventional expectation.
Rather than disrupting a dust population, storms create and concentrate one.

\subsection{Connection to Observed Magnetospheric Dust Swarms}

This trapping mechanism connects directly to an independently observed and currently unexplained phenomenon.
\citet{sommer2024} showed that clusters of sub-micron dust particles detected by the HEOS-2 and GORID instruments occur throughout the magnetosphere up to $\SI{60000}{km}$ altitude---and that these clusters are correlated with increased geomagnetic activity.

The proposed creation mechanism traces to \citet{horanyi1988}: fluffy meteoroids undergo electrostatic disruption when storm-enhanced plasma charging exceeds their tensile strength, producing swarms of fragments.
The storm-trapping mechanism described here provides the missing next step: the fragments become magnetically trapped, concentrating at $L \approx 6.6$.

\subsection{Dust Accumulation}

The micrometeoroid flux at 1~AU for grains larger than $\SI{1}{\micro m}$ is $F \sim \SI{e-4}{m^{-2}\,s^{-1}}$.
Assuming storms occupy ${\sim}15$\% of time and ${\sim}10$\% of trapped grains are retained, the steady-state density in the GEO trapping torus reaches $n_d \sim \SI{1.5}{m^{-3}}$ with sputtering as the dominant loss ($\tau \sim 10^5$~years).

Approximately half of the micrometeoroid flux at 1~AU is cometary.
Cometary dust contains water ice and hydrated minerals with albedos of 0.5--0.9, far higher than bare silicate (0.05--0.15).
Storm-fragmented cometary grains retain their ice content in the cold shadow environment at GEO (${\sim}80$--$\SI{100}{K}$), contributing high-albedo material to the trapped population.

\subsection{The Storm Cycle}

The mechanism produces the observed temporal profile:

\begin{enumerate}
\item \textbf{Storm (days 0--7):} Hot plasma fills GEO. Dust grains charge to high potentials and become magnetically trapped. Turbulent conditions prevent any ordered structures or stable specular reflection. Transient rate: \textit{suppressed}.

\item \textbf{Early recovery (days 7--21):} Storm subsides. Plasmasphere begins refilling. Conditions transitioning. Transient rate: \textit{still suppressed} (PSO-1: 55\% of baseline).

\item \textbf{Aggregation window (days 25--45):} Plasmasphere refilled. Cold, quiet conditions restored. Storm-concentrated dust population at elevated density. Favorable conditions for aggregation and specular reflection from icy dust structures. Transient rate: \textit{overshoot} (PSO-2: 309\% of baseline).

\item \textbf{Return to baseline (days 45+):} Concentrated dust disperses as grain charges decrease and gyroradii expand. Transient rate returns to equilibrium.
\end{enumerate}

\section{Dust Aggregation and Self-Organization}
\label{sec:aggregation}

Storm-trapped dust at GEO exists in a magnetically confined environment with low relative velocities and extended residence times.
Under these conditions, two processes---one well-established and one speculative---can produce the reflective structures required by the optical observations.

\subsection{Icy Cometary Dust: A Sticky Population}

Approximately half of the micrometeoroid flux at 1~AU is cometary in origin \citep{goertz1989}.
Cometary dust contains water ice and hydrated minerals, as confirmed by the Rosetta mission's in situ measurements at comet 67P/Churyumov-Gerasimenko, which found that the nucleus formed from ``gentle gravitational collapse of bound clumps of mm-sized dust aggregates'' intermixed with microscopic ice particles \citep{blum2017}.

Laboratory measurements by \citet{gundlach2015} demonstrated that micrometer-sized water-ice particles are approximately an order of magnitude stickier than silicate grains of comparable size, with higher surface energy and higher fragmentation threshold velocities.
In coagulation simulations, sub-micrometer icy particles grow directly to large aggregate sizes without encountering the bouncing barrier that limits silicate grain growth \citep{ormel2007,blum2018}.

Twenty-five years of laboratory research on dust agglomeration have established three collision regimes: ``hit-and-stick'' at low velocities, producing porous, fluffy aggregates; compaction at intermediate velocities; and fragmentation only at high velocities \citep{blum2018}.
At the thermal velocities expected for dust grains at radiative equilibrium ($T_d \sim \SI{300}{K}$), collision speeds are fractions of a mm\,s$^{-1}$---deep in the hit-and-stick regime.

\subsection{Aggregation in the Magnetic Trap}

Storm-trapped dust grains at GEO are magnetically confined with gyroradii of $0.5$--$\SI{5}{km}$ (Table~3).
Within a common trapping region, grains undergo repeated low-velocity encounters over weeks to months of residence time.
Each collision in the hit-and-stick regime produces a larger, more porous aggregate.

The resulting structures are expected to be anisotropic.
The magnetic field at GEO creates a preferred axis: grain motion perpendicular to $\mathbf{B}$ is confined by gyration while motion parallel to $\mathbf{B}$ is relatively free.
This anisotropy favors the formation of flattened, pancake-like aggregates---oblate structures with partially reflective icy surfaces.

Laboratory experiments in dusty plasmas confirm that charged grains self-organize into complex structures including chains, zigzags, and planar configurations \citep{goree2004,pnas2025}.
\citet{hartzell2013} observed that ``clumping is most often observed in small grains'' under electrostatic conditions.
Dust acoustic waves in laboratory plasmas lead to grain--grain collisions with subsequent growth of aggregate structures \citep{goree2004}.

A fluffy icy aggregate $1$--$\SI{4}{m}$ in diameter, with a partially reflective surface at $1$--$10\%$ specular efficiency, produces the observed plate magnitude at $\SI{42000}{km}$ (see Table~\ref{tab:degraded}).
This scenario requires no exotic physics---only the confinement of sticky icy dust in a magnetic bottle for a sufficient duration.

The empirical results (Part~I) do not constrain the internal structure of the aggregates.
They require only a storm-responsive, reflective, naturally occurring source at GEO altitude.
Whether the trapped dust population forms disordered fluffy aggregates, strongly coupled liquid-like structures, or more ordered configurations is a question for future laboratory and theoretical work under GEO-relevant plasma conditions.

\section{Specular Reflection Constraint from Flux Dilution}
\label{sec:specular}

The Solano et al.\ (2024) triple transient provides a quantitative test of the dust aggregate mechanism through the flux dilution argument (B.\ Villarroel, private communication).

\subsection{Instantaneous Magnitude}

A photographic plate integrates all photons received during the full exposure.
If a transient source is visible for duration $t_{\text{flash}}$ within an exposure of duration $t_{\text{exp}}$, the recorded flux is diluted by $D = t_{\text{flash}} / t_{\text{exp}}$.
The instantaneous magnitude is:
\begin{equation}
m_{\text{inst}} = m_{\text{plate}} + 2.5 \log_{10}\!\left(\frac{t_{\text{flash}}}{t_{\text{exp}}}\right)
\end{equation}

For the triple transient ($m_{\text{plate}} \approx 16$, $t_{\text{exp}} = \SI{3000}{s}$, $t_{\text{flash}} = \SI{0.5}{s}$):
\begin{equation}
m_{\text{inst}} = 16 + 2.5 \log_{10}\!\left(\frac{0.5}{3000}\right) = 16 - 9.4 = 6.6
\end{equation}

A source recorded at magnitude~16 on the plate was instantaneously magnitude~6.6---visible to the naked eye.

\subsection{Minimum Reflective Area: Lambertian vs Specular}

For a Lambertian reflector of cross-section $A$ and albedo $\alpha$ at distance $d$:
\begin{equation}
A_{\text{Lamb}} = \frac{\pi d^2 \times 10^{(m_\odot - m_{\text{inst}})/2.5}}{\alpha}
\end{equation}
where $m_\odot = -26.74$ (solar apparent magnitude, R band).
At $d = \SI{4.2e7}{m}$, $m_{\text{inst}} = 6.6$, and $\alpha = 0.5$ (icy cometary grains):
$A_{\text{Lamb}} \approx \SI{530}{m^2}$, corresponding to a diameter of ${\sim}\SI{26}{m}$.

For a flat specular reflector, sunlight is concentrated into a cone equal to the Sun's solid angle ($\Omega_\odot = \SI{6.8e-5}{sr}$), yielding a gain factor of $\pi / \Omega_\odot \approx 46{,}000$ over Lambertian.
The required area becomes:
\begin{equation}
A_{\text{spec}} = \frac{A_{\text{Lamb}}}{46{,}000} \approx \SI{0.012}{m^2} = \SI{115}{cm^2}
\end{equation}

At 100\% specular efficiency, a flat reflective surface only ${\sim}\SI{12}{cm}$ in diameter produces a naked-eye-brightness flash at $\SI{42000}{km}$.
Real aggregates will have lower efficiency; Table~\ref{tab:degraded} shows the required diameter across a range of efficiencies.
At $1$--$10\%$ efficiency, consistent with a partially reflective icy aggregate, the required diameter is $0.4$--$\SI{1.2}{m}$---well within the size range expected from weeks of hit-and-stick aggregation in a magnetic trap.

\subsection{Triple Transient Geometry}

Three sources within $\SI{10}{arcsec}$ at $d = \SI{42000}{km}$ are separated by ${\sim}\SI{2}{km}$.
Under warm plasmaspheric conditions ($T_e \sim 5$--$\SI{10}{eV}$), the gyroradius of a trapped $\SI{5}{\micro m}$ grain is $50$--$\SI{100}{km}$.
Three dust aggregates within $\SI{2}{km}$ are well inside a single magnetic trapping region---consistent with multiple aggregates forming within the same flux tube from storm-concentrated dust.

Two alternative interpretations are excluded.
First, a single object in motion would require a velocity of $\SI{0.68}{m\,s^{-1}}$---0.02\% of the GEO orbital velocity---to traverse $\SI{2}{km}$ during the 50-minute exposure.
An orbiting body at GEO cannot be this nearly stationary.
Second, a single tumbling object producing three spatially separated point sources would need to subtend $\SI{10}{arcsec}$ at $\SI{42000}{km}$, requiring a physical diameter of ${\sim}\SI{2}{km}$---far too large for any dust aggregate.
The three sources are therefore three independent structures.

\subsection{Flash Duration and Tumble Rate}

A tumbling specular reflector sweeps the solar reflection cone across the observer.
The flash duration is $t_{\text{flash}} = \theta_\odot / \omega_{\text{tumble}}$, where $\theta_\odot = \SI{9.3}{mrad}$ is the Sun's angular diameter.
A $\SI{0.5}{s}$ flash requires a tumble period of ${\sim}\SI{340}{s}$ (${\sim}\SI{5.6}{min}$ per revolution).
This is a physically reasonable rotation rate for a charged aggregate subject to weak radiation pressure and magnetic gradient torques in the plasmaspheric environment.

\subsection{Significance and Robustness}

The specular reflection model resolves the principal quantitative question: whether dust aggregates of plausible size can produce detectable optical signatures at $\SI{42000}{km}$.
The answer is affirmative.
At $1$--$10\%$ specular efficiency, an icy aggregate between $0.4$ and $\SI{1.2}{m}$ in diameter produces a flash brighter than magnitude~7 at GEO distance.

The result is robust against realistic degradation of the specular assumption.
A real icy aggregate will not be a perfect mirror: surface roughness, porosity, and partial disorder reduce the specular gain below the theoretical maximum of $\pi/\Omega_\odot \approx 46{,}000$.
Table~\ref{tab:degraded} shows the required aggregate diameter as a function of specular efficiency.

\begin{table}[H]
\centering
\caption{Required aggregate diameter under degraded specular efficiency ($t_{\text{flash}} = \SI{0.5}{s}$, $\alpha = 0.5$, $m_{\text{plate}} = 16$).}
\label{tab:degraded}
\begin{tabular}{@{}rrrrl@{}}
\toprule
Efficiency & Effective gain & Area ($\si{m^2}$) & Diameter & Status \\
\midrule
100\% & 46{,}222 & 0.012 & 12 cm & Viable \\
10\% & 4{,}622 & 0.12 & 38 cm & Viable \\
1\% & 462 & 1.15 & 1.2 m & Viable \\
0.1\% & 46 & 11.5 & 3.8 m & Viable \\
\bottomrule
\end{tabular}
\end{table}

Even at 1\% of theoretical efficiency---where 99\% of the aggregate surface is optically defective---the required diameter is only $\SI{1.2}{m}$.
At 0.1\%, the required diameter is $\SI{3.8}{m}$.
The model survives three orders of magnitude of degradation before the required sizes become physically implausible.
This margin ensures that the specular constraint is not sensitive to idealized assumptions about aggregate surface quality.

\section{The Extinction}
\label{sec:extinction}

The VASCO transients appear exclusively on pre-1957 plates and are absent in all modern surveys, including deep imaging to magnitude~25.5 with the Gran Telescopio Canarias at the location of the triple transient \citep{solano2024}.
The extinction has a plausible physical explanation involving three successive factors, each supported by published observations.

\subsection{Phase 1: Atmospheric Nuclear Testing (1945--1963)}

Over 500 atmospheric nuclear tests were conducted between 1945 and 1963.
High-altitude detonations injected energetic particles directly into the magnetosphere, and even surface tests produced electromagnetic pulses and ionospheric disturbances that propagated to magnetospheric altitudes.

The most extreme single event was Starfish Prime (July 9, 1962): a 1.4-megaton detonation at $\SI{400}{km}$ altitude that created an artificial radiation belt with electron fluxes over four orders of magnitude above natural levels \citep{stassinopoulos1971}.
The primary injection populated L-shells between approximately $1.2$ and $2.0$ \citep{stassinopoulos1971}, below the GEO trapping region at $L \approx 6.6$.
However, the enhanced trapped electron population drove intensified wave--particle interactions---whistler-mode chorus waves, enhanced pitch-angle scattering \citep{obrien2007}---that modified the electromagnetic environment across a broad range of L-shells, and DTRA assessments identified the event as a direct threat to geosynchronous satellites.

The cumulative effect of two decades of atmospheric testing was to keep the magnetosphere in a persistently disturbed state.
The storm-trapping model requires quiet plasmaspheric recovery phases lasting 25--45 days for dust aggregation to proceed (Section~\ref{sec:pso}, PSO-2).
Frequent artificial disturbances shortened or eliminated these quiet windows, degrading the conditions needed for aggregation.

The Partial Test Ban Treaty of August 1963 ended atmospheric testing---but in the same month, the first geosynchronous satellite was launched.

\subsection{Phase 2: Continuous Satellite Activity (1963--present)}

\citet{stephani2016} modeled spacecraft hydrazine thruster plume interactions with the magnetospheric plasma at GEO, demonstrating that each station-keeping burn injects combustion products (H$_2$, N$_2$, NH$_3$) that undergo charge exchange and photoionization, locally modifying plasma density and temperature.
GEO satellites perform station-keeping maneuvers approximately every two weeks.
With hundreds of satellites currently at GEO, the orbital environment is subject to continuous, overlapping plasma disturbances.

Modern electric propulsion systems compound this effect.
Hall-effect thrusters produce xenon ion beams at energies of $100$--$\SI{1600}{eV}$, sufficient to sputter solid surfaces \citep{goebel2008}.
Charge-exchange ions from these beams spread over a much wider volume than the primary beam, creating extended regions of disturbed plasma.

The cumulative effect of continuous station-keeping, RF broadcasting from communications payloads, outgassing, and debris generation is to transform the GEO electromagnetic environment from its pristine pre-1963 state into a permanently disturbed one.
The quiet conditions required for slow dust aggregation over weeks-long recovery phases (Section~\ref{sec:aggregation}) cannot restore in this environment.

\subsection{Phase 3: Modern Survey Pipeline Rejection}

The absence of VASCO-type transients in modern CCD surveys does not independently confirm extinction, because modern survey reduction pipelines systematically reject single-exposure detections as instrumental artifacts.
A sub-second specular flash from GEO appearing on a $\SI{60}{s}$ CCD exposure would be recorded at approximately magnitude~22---faint, unresolved, present in one frame, and absent in the next.
This signature is indistinguishable from a cosmic ray hit.
Automated artifact rejection removes it before it reaches the catalog.

The POSS-I photographic plates detected these transients precisely because emulsion-based recording has no automated rejection.
Every photon that reached the plate was recorded.
The ``primitive'' technology was, for this specific phenomenon, the more sensitive detector.

A dedicated search for sub-second GEO-altitude transients in archival CCD data, with single-frame artifact rejection disabled for unresolved point sources, would constitute a direct test of whether any residual population survives.

\subsection{A Testable Prediction}

The three-phase model makes a specific prediction.
Photographic plates taken between 1958 and mid-1962---after the end of POSS-I but before the most intense phase of atmospheric nuclear testing and the onset of GEO satellite activity---should still show VASCO-type transients, as the GEO environment remained largely pristine during this window.
Plates taken after mid-1962 should show a decline, and plates from the late 1960s onward should show complete absence.
European observatory archives in the APPLAUSE database contain plates from this period (Section~\ref{sec:limitations}) and could be used to test this prediction if transient detection is performed.

\begin{figure}[H]
\centering
\includegraphics[width=\textwidth]{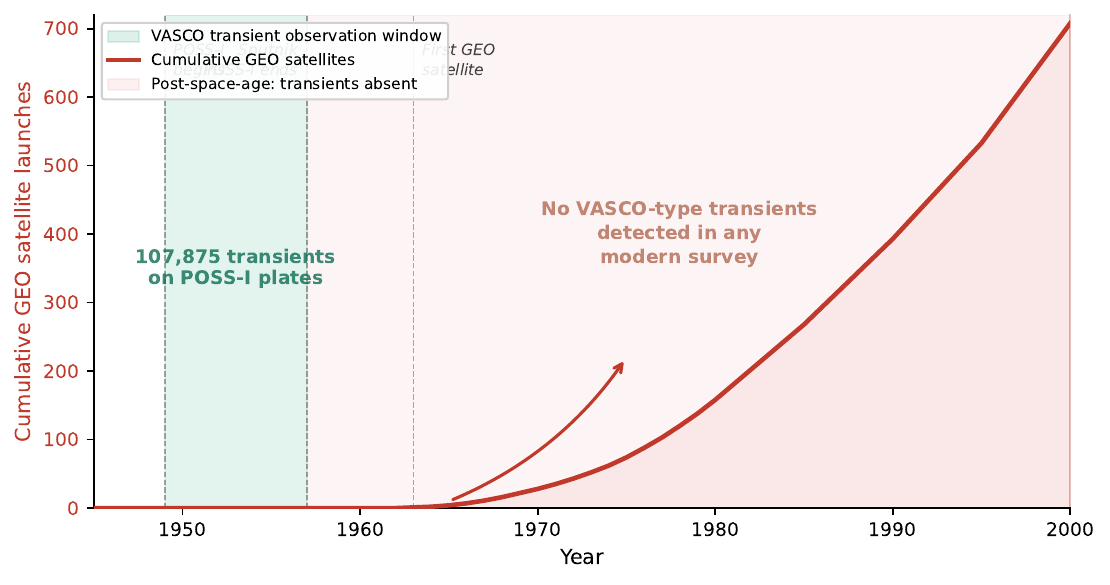}
\caption{Temporal relationship between VASCO transient observations and the growth of the GEO satellite population. The green shaded region marks the POSS-I observation window during which 107,875 transients were detected. The red curve shows the cumulative number of satellites launched into geosynchronous orbit. Atmospheric nuclear testing (1945--1963) and particularly the Starfish Prime detonation (July 1962) disrupted the magnetospheric environment during the transition period. No VASCO-type transients have been detected in any survey conducted after the onset of artificial GEO activity.}
\label{fig:extinction}
\end{figure}

Figure~\ref{fig:extinction} shows the temporal relationship between the VASCO transient observation window and the subsequent growth of the GEO satellite population.

\section{Observational Constraints Satisfied}
\label{sec:constraints}

\begin{enumerate}
\item \textbf{Dose-dependent Kp suppression}: Satisfied. Storms disrupt the source environment.
\item \textbf{Geosynchronous altitude}: Satisfied. $L \approx 6.6$.
\item \textbf{Earth-shadow deficit}: Satisfied. No sunlight in shadow, no specular reflection.
\item \textbf{Point-source PSF}: Satisfied. Compact aggregate at $\SI{42000}{km}$ is unresolved.
\item \textbf{Sub-second duration}: Satisfied. Specular geometry from tumbling aggregate is transient.
\item \textbf{Nuclear test enhancement}: Consistent. Independent atmospheric excitation mechanism.
\item \textbf{Post-storm suppression (days 7--21)}: Satisfied. Plasmasphere not yet refilled.
\item \textbf{Post-storm overshoot (days 25--45)}: Satisfied. Storm-concentrated material under restored conditions.
\item \textbf{Pre-Sputnik occurrence}: Satisfied. Natural process.
\item \textbf{Post-1957 extinction}: Satisfied. Three-phase disruption: Starfish Prime artificial belt (1962), continuous satellite thruster plume disturbance (1963--present), and modern survey pipeline rejection of single-frame detections (Section~\ref{sec:extinction}).
\item \textbf{Storm-correlated dust swarms}: Explained. Connects to \citet{sommer2024}.
\item \textbf{Specular size constraint}: Satisfied. An icy aggregate $1$--$\SI{4}{m}$ in diameter at $1$--$10\%$ specular efficiency produces the observed plate magnitude at $\SI{42000}{km}$ (Section~\ref{sec:specular}).
\end{enumerate}

\section{Predictions Requiring Further Testing}
\label{sec:predictions}

\begin{enumerate}
\item \textbf{Multi-site replication.} The Kp staircase and the days 25--45 overshoot should replicate at every observatory with pre-Sputnik plate archives. Pre-registered predictions exist for Hamburg and Potsdam. This is the most important next step.

\item \textbf{Stellar detection rate test.} Star counts should be flat across storm and quiet periods. If transient counts vary while star counts do not, the effect is source-specific.

\item \textbf{Transient extinction timeline.} Transient rates should decline in correlation with cumulative GEO object count, not at any single date.

\item \textbf{Full recovery curve.} Sliding-window analysis from day~0 through day~60 should map the complete suppression--overshoot--return profile.

\item \textbf{Nuclear enhancement absent at European sites.} Sites $>$\SI{5000}{km} from test locations should not show the prompt nuclear correlation.

\item \textbf{Spectral diagnostics.} If transient spectra can be extracted, dusty plasma aggregates should show both continuum reflection and plasma emission features.

\item \textbf{Exposure-duration magnitude correlation.} If the transients are sub-second specular flashes, their apparent plate magnitude should correlate with exposure duration across different plate archives: shorter exposures should record brighter apparent magnitudes for the same source population, as predicted by the flux dilution framework (Section~\ref{sec:specular}).
\end{enumerate}

\section{Discussion}
\label{sec:discussion}

\subsection{Relationship to Villarroel's Interpretation}

The VASCO team interprets the transients as possible reflections from artificial objects of unknown origin in pre-Sputnik orbit.
The storm-trapping mechanism proposed here provides a natural alternative that satisfies all the same observational constraints.

Crucially, the present work \textit{validates} the VASCO team's core observational claims: the transients are real, they are at GEO altitude, and they have reflective surfaces.
The dose--response staircase and the post-storm overshoot provide the strongest evidence to date that these claims are correct.
The disagreement concerns only the \textit{nature} of the reflective source, not its existence.

\subsection{Why These Connections Were Not Made Earlier}

The storm-trapping mechanism required connecting five separate research domains that have had minimal cross-fertilization:

\begin{enumerate}
\item VASCO photographic plate transients \citep{villarroel2020}
\item The Kp dose--response (Cann 2026)
\item Laboratory dusty plasma self-organization \citep{goree2004}
\item Magnetospheric dust dynamics \citep{horanyi1988,goertz1989}
\item Storm-correlated magnetospheric dust swarms \citep{sommer2024}
\end{enumerate}

No individual research community had reason to connect these observations.
The Kp dose--response, established only in 2026, provided the critical empirical bridge.

\subsection{Limitations}
\label{sec:limitations}

The principal limitations of this work are:

\begin{enumerate}
\item The PSO-2 overshoot test was motivated by the PSO-1 null result. While pre-registered before analysis and robust across all checks, the window selection was informed by prior data from the same dataset.

\item Only 6 storms met the 25-day isolation criterion, yielding 126 recovery dates. The effect is statistically significant but based on a small number of independent storm events.

\item The dust aggregation mechanism remains theoretical. No in situ observations of storm-trapped dust aggregates at GEO have been reported.

\item All results derive from a single observatory (Palomar). Multi-site replication is essential.
\end{enumerate}

\subsection{Multi-Site Replication Feasibility}

An analysis of photographic plate archives at five European observatories in the APPLAUSE DR4 database confirms that replication is feasible.
Hamburg (7{,}509 plates), Tartu (1{,}360), Bamberg (1{,}565), Vatican (543), and Potsdam (383) all have plates spanning the VASCO-relevant period 1949--1957.
A chi-squared test of observing rates during storm periods (Kp $\geq 5$) versus quiet periods (Kp $< 3$) shows no significant difference at any site (all $p > 0.05$; Tartu: ratio 0.993, $p = 1.00$; Hamburg: ratio 1.032, $p = 0.68$).
The 1950s astronomers at these observatories observed with complete indifference to geomagnetic conditions.
This establishes that the APPLAUSE plate archives are unbiased with respect to Kp and that any future storm-dependent transient detection results at these sites will reflect the physical response of the source, not an artifact of the observing schedule.

\section{Conclusions}
\label{sec:conclusions}

The VASCO photographic plate transients are real, magnetospherically coupled, reflective phenomena at geosynchronous altitude.
The empirical evidence---a dose--response staircase at $3.2\sigma$ and a post-storm overshoot at $3.2\sigma$, combining to $3.6$--$4.7\sigma$---establishes a storm-driven suppression--overshoot--return temporal profile that requires an active concentration mechanism.
Regardless of the proposed mechanism, this temporal profile constitutes a new observational constraint that any explanation of the VASCO transients must satisfy.

Storm-enhanced electromagnetic trapping of charged micrometeoroid dust provides a candidate mechanism satisfying all twelve known observational constraints, including the post-storm overshoot and the extinction of transients following the onset of the space age.
Aggregation of the trapped icy dust into partially reflective structures represents the candidate mechanism proposed within this framework.
A flux dilution analysis of the triple transient reported by \citet{solano2024} demonstrates that specular reflection from an icy aggregate of only $1$--$\SI{4}{m}$ diameter at $1$--$10\%$ efficiency produces the observed plate magnitude at $\SI{42000}{km}$.

These results are substantial but require independent confirmation.
Multi-site replication using Hamburg and Potsdam plate archives, with the overshoot test pre-registered at the days 25--45 window, would provide the definitive test.
If confirmed, the VASCO transients may represent the first and last observations of a natural magnetospheric phenomenon that was inadvertently destroyed by the beginning of the space age.

\section*{Acknowledgments}

The author gratefully acknowledges Dr.\ Beatriz Villarroel for establishing the VASCO project, for her persistent insistence that the photographic plate transients represent real astrophysical phenomena---a conclusion that the work presented here has confirmed at $3.6$--$4.7\sigma$ combined significance---and for the flux dilution insight that enabled the specular reflection constraint (Section~\ref{sec:specular}).
Her observational foundation made the present work possible.

Stephen Bruehl is acknowledged for sharing the VASCO transient dataset.

Brian Doherty is acknowledged for independent replication of the shadow deficit and nuclear test correlation analyses.

This work was conducted independently using personal computing resources with no institutional or external funding.

\section*{Data Availability}

Data and code are archived at \url{https://osf.io/u9nas}.
This repository includes reproduction scripts for the post-storm overshoot analysis and the flux dilution test, pre-registered predictions for multi-site replication at Hamburg and Potsdam, and observing schedule independence results for five European observatories.
The transient dataset is from the supplementary materials of \citet{bruehl2025} (DOI: 10.1038/s41598-025-21620-3).
The Kp archive is from GFZ Potsdam (\url{https://kp.gfz.de/app/files/Kp_ap_since_1932.txt}, CC BY 4.0).

\end{document}